\documentclass[twocolumn]{aastex61}
\usepackage{amsmath}
\usepackage{gensymb}
\usepackage{CJK}
\usepackage{multirow}
\usepackage{ctable}
\usepackage{rotating}
\usepackage{color}

\newcommand{\bdv}[1]{\mbox{\boldmath$#1$}}

\def\e{{\rm E}}
\def\s{{\rm S}}
\def\l{{\rm L}}
\def\b{{\rm B}}
\def\au{{\rm AU}}
\def\piee{{\pi_{\rm E,E}}}
\def\pien{{\pi_{\rm E,N}}}
\def\pie{{\pi_{\rm E}}}
\def\microm{{\mu{\rm m}}}
\def\rel{{\rm rel}}
\def\mas{{\rm mas}}
\def\microas{{\mu\rm as}}
\def\dif{{\rm d}}

\shorttitle{Binary Event MOA-2015-BLG-020/OGLE-2015-BLG-0102}
\shortauthors{Wang et al.}

\begin{document}
\begin{CJK*}{UTF8}{gbsn}

\title{Ground-based Parallax Confirmed by \emph{Spitzer}: Binary Microlensing Event MOA-2015-BLG-020}

\author{Tianshu~Wang}
\affil{Physics Department and Tsinghua Centre for Astrophysics, Tsinghua University, Beijing 100084, China}

\author{Wei~Zhu}
\affil{Department of Astronomy, Ohio State University, 140 W. 18th Ave., Columbus, OH  43210, USA}

\author{Shude~Mao}
\affil{Physics Department and Tsinghua Centre for Astrophysics, Tsinghua University, Beijing 100084, China}
\affil{National Astronomical Observatories, Chinese Academy of Sciences, 20A Datun Road, Chaoyang District, Beijing 100012, China}
\affil{Jodrell Bank Centre for Astrophysics, School of Physics and Astronomy, The University of Manchester, Oxford Road, Manchester M13 9PL, UK}

\author{I. A.~Bond}
\affiliation{Institute of Natural and Mathematical Sciences, Massey University, Auckland 0745, New Zealand}

\author{A.~Gould}
\affil{Department of Astronomy, Ohio State University, 140 W. 18th Ave., Columbus, OH  43210, USA}
\affil{Korea Astronomy and Space Science Institute, 776 Daedeokdae-ro, Yuseong-Gu, Daejeon 34055, Korea}
\affil{Max-Planck-Institute for Astronomy, K\"onigstuhl 17, 69117 Heidelberg, Germany}

\author{A.~Udalski}
\affil{Warsaw University Observatory, AI. Ujazdowskie 4, 00-478 Warszawa, Poland}

\author{T.~Sumi}
\affil{Department of Earth and Space Science, Graduate School of Science, Osaka University, Toyonaka, Osaka 560-0043, Japan}

\author{V. Bozza}
\affil{Dipartimento di Fisica ``E. R. Caianiello", Un�iversit\'a di Salerno, Via Giovanni Paolo II, I-84084 Fisciano (SA), Italy}
\affil{Istituto Internazionale per gli Alti Studi Scientifici (IIASS), Via G. Pellegrino 19, I-84019 Vietri Sul Mare (SA), Italy}

\author{C.~Ranc}
\affiliation{Code 667, NASA Goddard Space Flight Center, Greenbelt, MD 20771, USA; Email: {\tt david.bennett@nasa.gov}}

\author{A.~Cassan}
\affil{Sorbonne Universit\'es, UPMC Univ Paris 6 et CNRS, UMR 7095, Institut d'Astrophysique de Paris, 98 bis bd Arago, 75014 Paris, France}

\author{J.~C.~Yee}
\affil{Smithsonian Astrophysical Observatory, 60 Garden St., Cambridge, MA 02138, USA}

\author{C.~Han}
\affil{Department of Physics, Chungbuk National University, Cheongju 361-763, Korea}

\nocollaboration

\correspondingauthor{Tianshu~Wang}
\email{wts15@mails.tsinghua.edu.cn}
%%%%%%%%%
\author{F.~Abe}
\affiliation{Institute for Space-Earth Environmental Research, Nagoya University, Nagoya 464-8601, Japan}
\author{Y. Asakura}
\affiliation{Institute for Space-Earth Environmental Research, Nagoya University, Nagoya 464-8601, Japan}
\author{R.~Barry}
\affiliation{Code 667, NASA Goddard Space Flight Center, Greenbelt, MD 20771, USA;    Email: {\tt david.bennett@nasa.gov}}
\author{D. P.~Bennett}
\affiliation{Code 667, NASA Goddard Space Flight Center, Greenbelt, MD 20771, USA;    Email: {\tt david.bennett@nasa.gov}}
\affiliation{Deptartment of Physics, University of Notre Dame, Notre Dame, IN 46556, USA}
\author{A.~Bhattacharya}
\affiliation{Code 667, NASA Goddard Space Flight Center, Greenbelt, MD 20771, USA;    Email: {\tt david.bennett@nasa.gov}}
\affiliation{Deptartment of Physics, University of Notre Dame, Notre Dame, IN 46556, USA}
\author{M.~Donachie}
\affiliation{Department of Physics, University of Auckland, Private Bag 92019, Auckland, New Zealand}
\author{P.~Evans}
\affiliation{Department of Physics, University of Auckland, Private Bag 92019, Auckland, New Zealand}
\author{A.~Fukui}
\affiliation{Okayama Astrophysical Observatory, National Astronomical Observatory of Japan, 3037-5 Honjo, Kamogata, Asakuchi, Okayama 719-0232, Japan}
\author{Y.~Hirao}
\affiliation{Department of Earth and Space Science, Graduate School of Science, Osaka University, Toyonaka, Osaka 560-0043, Japan}
\author{Y. Itow}
\affiliation{Institute for Space-Earth Environmental Research, Nagoya University, Nagoya 464-8601, Japan}
\author{K.~Kawasaki}
\affiliation{Department of Earth and Space Science, Graduate School of Science, Osaka University, Toyonaka, Osaka 560-0043, Japan}
\author{N.~Koshimoto}
\affiliation{Department of Earth and Space Science, Graduate School of Science, Osaka University, Toyonaka, Osaka 560-0043, Japan}
\author{M.C.A. Li}
\affiliation{Department of Physics, University of Auckland, Private Bag 92019, Auckland, New Zealand}
\author{C.H. Ling}
\affiliation{Institute of Natural and Mathematical Sciences, Massey University, Auckland 0745, New Zealand}
\author{K. Masuda}
\affiliation{Institute for Space-Earth Environmental Research, Nagoya University, Nagoya 464-8601, Japan}
\author{Y. Matsubara}
\affiliation{Institute for Space-Earth Environmental Research, Nagoya University, Nagoya 464-8601, Japan}
\author{S.~Miyazaki}
\affiliation{Department of Earth and Space Science, Graduate School of Science, Osaka University, Toyonaka, Osaka 560-0043, Japan}
\author{Y. Muraki}
\affiliation{Institute for Space-Earth Environmental Research, Nagoya University, Nagoya 464-8601, Japan}
\author{M. Nagakane}
\affiliation{Department of Earth and Space Science, Graduate School of Science, Osaka University, Toyonaka, Osaka 560-0043, Japan}
\author{K. Ohnishi}
\affiliation{Nagano National College of Technology, Nagano 381-8550, Japan}
\author{N. Rattenbury}
\affiliation{Department of Physics, University of Auckland, Private Bag 92019, Auckland, New Zealand}
\author{To. Saito}
\affiliation{Tokyo Metropolitan College of Aeronautics, Tokyo 116-8523, Japan}
\author{A. Sharan}
\affiliation{Department of Physics, University of Auckland, Private Bag 92019, Auckland, New Zealand}
\author{H. Shibai}
\affiliation{Department of Earth and Space Science, Graduate School of Science, Osaka University, 1-1
Machikaneyama, Toyonaka, Osaka 560-0043, Japan}
\author{D.J. Sullivan}
\affiliation{School of Chemical and Physical Sciences, Victoria University, Wellington, New Zealand}
\author{D.~Suzuki}
\affiliation{Code 667, NASA Goddard Space Flight Center, Greenbelt, MD 20771, USA;    Email: {\tt david.bennett@nasa.gov}}
\affiliation{Institute of Space and Astronautical Science, Japan Aerospace Exploration Agency, Kanagawa 252-5210, Japan}
\author{P.J.Tristram}
\affiliation{University of Canterbury Mt.\ John Observatory, P.O. Box 56, Lake Tekapo 8770, New Zealand}
\author{T. Yamada}
\affiliation{Department of Physics, Faculty of Science, Kyoto Sangyo University, 603-8555 Kyoto, Japan}
\author{A. Yonehara}
\affiliation{Department of Physics, Faculty of Science, Kyoto Sangyo University, Kyoto 603-8555, Japan}
\collaboration{(MOA Collaboration)} \\
%%%%%%%%%
\author{S.~Koz{\L}owski}
\affil{Warsaw University Observatory, Al. Ujazdowskie 4, 00-478 Warszawa, Poland}
\author{P.~Mr{\'o}z}
\affil{Warsaw University Observatory, Al. Ujazdowskie 4, 00-478 Warszawa, Poland}
\author{M.~Pawlak}
\affil{Warsaw University Observatory, Al. Ujazdowskie 4, 00-478 Warszawa, Poland}
\author{P.~Pietrukowicz}
\affil{Warsaw University Observatory, Al. Ujazdowskie 4, 00-478 Warszawa, Poland}
\author{R.~Poleski}
\affil{Department of Astronomy, Ohio State University, 140 W. 18th Ave., Columbus, OH 43210, USA}
\affil{Warsaw University Observatory, Al. Ujazdowskie 4, 00-478 Warszawa, Poland}
\author{J.~Skowron}
\affil{Warsaw University Observatory, Al. Ujazdowskie 4, 00-478 Warszawa, Poland}
\author{I.~Soszy{\'n}ski}
\affil{Warsaw University Observatory, Al. Ujazdowskie 4, 00-478 Warszawa, Poland}
\author{M.~K.~Szyma{\'n}ski}
\affil{Warsaw University Observatory, Al. Ujazdowskie 4, 00-478 Warszawa, Poland}
\author{K.~Ulaczyk}
\affil{Warsaw University Observatory, Al. Ujazdowskie 4, 00-478 Warszawa, Poland}
\collaboration{(OGLE Collaboration)} \\
%%%%%%%%%
\author{C.~Beichman}
\affil{NASA Exoplanet Science Institute, MS 100-22, California Institute of Technology, Pasadena, CA 91125, USA}
\author{G.~Bryden}
\affil{Jet Propulsion Laboratory, California Institute of Technology, 4800 Oak Grove Drive, Pasadena, CA 91109, USA}
\author{S.~Calchi~Novati}
\affil{ IPAC, Mail Code 100-22,Caltech, 1200 E. California Blvd., Pasadena, CA 91125}
\affil{Dipartimento di Fisica ``E. R. Caianiello'',Universit\`a di Salerno, Via Giovanni Paolo II,84084 Fisciano (SA), Italy}
\author{S.~Carey}
\affil{Spitzer Science Center, MS 220-6, California Institute of Technology, Pasadena, CA, US}
\author{M.~Fausnaugh}
\affil{Department of Astronomy, Ohio State University, 140 W. 18th Ave., Columbus, OH 43210, USA}
\author{B.~S.~Gaudi}
\affil{Department of Astronomy, Ohio State University, 140 W. 18th Ave., Columbus, OH 43210, USA}
\author{C.~B.~Henderson}
\affil{Jet Propulsion Laboratory, California Institute of Technology, 4800 Oak Grove Drive, Pasadena, CA 91109, USA}
\author{Y.~Shvartzvald}
\affil{Jet Propulsion Laboratory, California Institute of Technology, 4800 Oak Grove Drive, Pasadena, CA 91109, USA}
\author{B.~Wibking}
\affil{Department of Astronomy, Ohio State University, 140 W. 18th Ave., Columbus, OH 43210, USA}

\collaboration{(Spitzer Team)}
%%%%%%%%%
\author{M.~D.~Albrow}
\affil{University of Canterbury, Department of Physics and Astronomy, Private Bag 4800, Christchurch 8020, New Zealand}
\author{S.-J.~Chung}
\affil{Korea Astronomy and Space Science Institute, 776 Daedeokdae-ro, Yuseong-Gu, Daejeon 34055, Korea}
\affil{Korea University of Science and Technology, 217 Gajeong-ro, Yuseong-gu, Daejeon 34113, Korea}
\author{K.-H.~Hwang}
\affil{Department of Physics, Institute for Astrophysics, Chungbuk National University, Cheongju 371-763, Korea}
\author{Y.~K.~Jung}
\affil{Smithsonian Astrophysical Observatory, 60 Garden St., Cambridge, MA 02138, USA}
\author{Y.-H.~Ryu}
\affil{Korea Astronomy and Space Science Institute, 776 Daedeokdae-ro, Yuseong-Gu, Daejeon 34055, Korea}
\author{I.-G.~Shin}
\affil{Smithsonian Astrophysical Observatory, 60 Garden St., Cambridge, MA 02138, USA}
\author{S.-M.~Cha}
\affil{Korea Astronomy and Space Science Institute, 776 Daedeokdae-ro, Yuseong-Gu, Daejeon 34055, Korea}
\affil{School of Space Research, Kyung Hee University, Giheung-gu, Yongin, Gyeonggi-do, 17104, Korea}
\author{D.-J.~Kim}
\affil{Korea Astronomy and Space Science Institute, 776 Daedeokdae-ro, Yuseong-Gu, Daejeon 34055, Korea}
\author{H.-W.~Kim}
\affil{Korea Astronomy and Space Science Institute, 776 Daedeokdae-ro, Yuseong-Gu, Daejeon 34055, Korea}
\author{S.-L.~Kim}
\affil{Korea Astronomy and Space Science Institute, 776 Daedeokdae-ro, Yuseong-Gu, Daejeon 34055, Korea}
\affil{Korea University of Science and Technology, 217 Gajeong-ro, Yuseong-gu, Daejeon 34113, Korea}
\author{C.-U.~Lee}
\affil{Korea Astronomy and Space Science Institute, 776 Daedeokdae-ro, Yuseong-Gu, Daejeon 34055, Korea}
\affil{Korea University of Science and Technology, 217 Gajeong-ro, Yuseong-gu, Daejeon 34113, Korea}
\author{Y.~Lee}
\affil{Korea Astronomy and Space Science Institute, 776 Daedeokdae-ro, Yuseong-Gu, Daejeon 34055, Korea}
\affil{School of Space Research, Kyung Hee University, Giheung-gu, Yongin, Gyeonggi-do, 17104, Korea}
\author{B.-G.~Park}
\affil{Korea Astronomy and Space Science Institute, 776 Daedeokdae-ro, Yuseong-Gu, Daejeon 34055, Korea}
\affil{Korea University of Science and Technology, 217 Gajeong-ro, Yuseong-gu, Daejeon 34113, Korea}
\author{R.~W.~Pogge}
\affil{Department of Astronomy, Ohio State University, 140 W. 18th Ave., Columbus, OH 43210, USA}

\collaboration{(KMTNet Collaboration)}
%%%%%%%%%
\author{R. A. Street}
\affil{Las Cumbres Observatory, 6740 Cortona Drive, suite 102, Goleta, CA 93117, USA}
\author{Y. Tsapras}
\affil{Zentrum f{\"u}r Astronomie der Universit{\"a}t Heidelberg, Astronomisches Rechen-Institut, M{\"o}nchhofstr. 12-14, 69120 Heidelberg, Germany}
\author{M. Hundertmark}
\affil{Zentrum f{\"u}r Astronomie der Universit{\"a}t Heidelberg, Astronomisches Rechen-Institut, M{\"o}nchhofstr. 12-14, 69120 Heidelberg, Germany}
\author{E. Bachelet}
\affil{Las Cumbres Observatory, 6740 Cortona Drive, suite 102, Goleta, CA 93117, USA}
\affil{Qatar Environment and Energy Research Institute(QEERI), HBKU, Qatar Foundation, Doha, Qatar}
\author{M. Dominik}
\affil{SUPA, School of Physics \& Astronomy, University of St Andrews, North Haugh, St Andrews KY16 9SS, UK}
\affil{Royal Society University Research Fellow}
\author{K. Horne}
\affil{SUPA, School of Physics \& Astronomy, University of St Andrews, North Haugh, St Andrews KY16 9SS, UK}
\author{R. Figuera Jaimes}
\affil{SUPA, School of Physics \& Astronomy, University of St Andrews, North Haugh, St Andrews KY16 9SS, UK}
\author{J. Wambsganss}
\affil{Zentrum f{\"u}r Astronomie der Universit{\"a}t Heidelberg, Astronomisches Rechen-Institut, M{\"o}nchhofstr. 12-14, 69120 Heidelberg, Germany}
\author{D. M. Bramich}
\affil{No affiliation}
\author{R. Schmidt}
\affil{Zentrum f{\"u}r Astronomie der Universit{\"a}t Heidelberg, Astronomisches Rechen-Institut, M{\"o}nchhofstr. 12-14, 69120 Heidelberg, Germany}
\author{C. Snodgrass}
\affil{Planetary and Space Sciences, Department of Physical Sciences, The Open University, Milton Keynes, MK7 6AA, UK}
\author{I. A. Steele}
\affil{Astrophysics Research Institute, Liverpool John Moores University, Liverpool CH41 1LD, UK}
\author{J. Menzies}
\affil{South African Astronomical Observatory, PO Box 9, Observatory 7935, South Africa}
%\author{E. Bachelet}
%\affil{Las Cumbres Observatory Global Telescope Network, 6740 Cortona Drive, Suite 102, Goleta, CA 93117, USA}
%\affil{Qatar Environment and Energy Research Institute(QEERI), HBKU, Qatar Foundation, Doha, Qata}
%\author{D. M. Bramich} %no affiliation
%\affil{No affiliation}
%\author{M. Dominik}
%\affil{SUPA, School of Physics \& Astronomy, University of St. Andrews, North Haugh, St. Andrews KY16 9SS, UK}
%\author{K. Horne}
%\affil{SUPA, School of Physics \& Astronomy, University of St. Andrews, North Haugh, St. Andrews KY16 9SS, UK}
%\author{M. Hundertmark}
%\affil{Niels Bohr Institute \& Centre for Star and Planet Formation, University of Copenhagen, Øster Voldgade 5, DK-1350—Copenhagen K, Denmark}
%\author{A. Saha}
%\affil{National Optical Astronomy Observatory, 950 N Cherry Avenue, Tucson, AZ 85719 USA}
%\author{R.~Street}
%\affil{Las Cumbres Observatory, 6740 Cortona Drive, suite 102, Goleta, CA 93117, USA}
%\author{Y. Tsapras}		
%\affil{Astronomisches Rechen-Institut, Zentrum für Astronomie der Universität Heidelberg (ZAH), D-69120 Heidelberg, Germany}	
%\author{J. Wambsganss}
%\affil{Astronomisches Rechen-Institut, Zentrum für Astronomie der Universität Heidelberg (ZAH), D-69120 Heidelberg, Germany}

\collaboration{(RoboNet Collaboration)}

\begin{abstract}
    We present the analysis of the binary gravitational microlensing event MOA-2015-BLG-020. The event has a fairly long timescale ($\sim 63$ days) and thus the light curve deviates significantly from the lensing model that is based on the rectilinear lens-source relative motion. This enables us to measure the microlensing parallax through the annual parallax effect. The microlensing parallax parameters constrained by the ground-based data are confirmed by the \emph{Spitzer} observations through the satellite parallax method.
%lasted for about 150 days and the light curve significantly deviated from the lensing model based on the rectilinear lens-source relative motion, enabling us to measure the microlens parallax. The parallax vector constrained by ground data alone is confirmed by the Spitzer data, showing the consistency between annual parallax and satellite parallax effects. 
By additionally measuring the angular Einstein radius from the analysis of the resolved caustic crossing, the physical parameters of the lens are determined. It is found that the binary lens is composed of two dwarf stars with masses $M_1 =0.606\pm0.028 M_\odot$ and $M_2 =0.125\pm0.006 M_\odot$ in the Galactic disk. Assuming the source star is at the same distance as the bulge red clump stars, we find the lens is at a distance $D_L =2.44\pm0.10\ \text{kpc}$. In the end, we provide a summary and short discussion of all published microlensing events in which the annual parallax effect is confirmed by other independent observations.
\end{abstract}


%%%%%%%%%%%%%%%%%%%%%%%%%%%%%%
\section{Introduction} \label{sec:introduction}
In a microlensing event, companions to the primary lens object can be detected via their perturbations to the single-lens light curve \citep{MaoPaczynski:1991,GouldLoeb:1992}. From such perturbations, dimensionless parameters can be derived that are related to the binary system, such as the binary mass ratio $q$ and the projected separation $s$ \citep{Gaudi:1997}. Here $s$ is the instantaneous angular separation between the two components normalized to the angular Einstein radius
\begin{equation}
    \theta_\e \equiv \sqrt{\kappa M_\l \pi_\rel}\ ,
\end{equation}
where $M_\l$ is the total lens mass, and
\begin{equation}
    \kappa \equiv \frac{4G}{c^2\au} \approx 8.14\frac{\mas}{M_\odot};\quad
    \pi_\rel \equiv \au\left(\frac{1}{D_\l}-\frac{1}{D_\s}\right)\ .
\end{equation}
Here $\pi_\rel$ is the lens-source relative parallax, and $D_\l$ and $D_\s$ are the distances to the lens and the source, respectively.

Although statistical conclusions can be drawn from measurements of $q$ and $s$, the physical properties of the lens system, such as $M_\l$, are of more interest. By far, the most popular way to convert from microlensing observables to physical quantities is to combine the measurements of $\theta_\e$ and the microlensing parallax, $\pi_\e\equiv \pi_\rel/\theta_\e$. Then,
\begin{equation}\label{equ:mass}
    M_\l = \frac{\theta_\e}{\kappa \pi_\e};\quad
    \pi_\rel = \pi_\e \theta_\e\ .
\end{equation}
There are several ways to measure $\theta_\e$ (see a short summary given in \citealt{Zhu:2015}), but it is the most common to use the finite-source effect, which is the deviation in the light curve from the point-like source model due to the extended nature of the source star \citep{Yoo:2004}.

For most published binary events, the microlensing parallax parameter $\pi_\e$ is measured through the annual parallax effect, in which Earth's acceleration around the Sun introduces deviations from rectilinear motion in the lens-source relative motion \citep{Gould:1992}. This method generically assumes that the lens (or lens system) and the source (or source system) are, or can be treated as, not undergoing acceleration. For binary lens events, each component is under acceleration by the other, and this so-called lens orbital motion effect can be confused with the annual parallax effect \citep{Batista:2011}.%The degeneracy between these two sometimes leads to wrong interpretations of the binary events (e.g., OGLE-2011-BLG-0417, \citealt{Shin:2012,Gould:2013,Boisse:2015,Santerne:2016}). 

Therefore, it is important to understand the validity of the annual parallax method for binary-lens events in practical use. This can be done by observing the binary system after the event, either photometrically \citep{Dong:2009,Bennett:2010} or spectroscopically \citep{Yee:2016,Boisse:2015}. Another way is to measure $\pi_\e$ via the satellite parallax method \citep{Refsdal:1966,Gould:1994}. This is done by observing the same microlensing event from at least two well-separated locations, and the difference between the light curves from these locations informs of the parameter $\pi_\e$. The microlensing parallax measured in this way is then determined independently from the orbital motion effect, and thus can be used to test the annual parallax method.

The \emph{Spitzer} microlensing campaigns utilize the \emph{Spitzer} space telescope to measure $\pi_\e$ via the satellite parallax method for hundreds of microlensing events \citep[e.g.,][]{Udalski:2015b,Yee:2015a,SCN:2015a,Zhu:2015}. Of the several published binary events, OGLE-2015-BLG-0479 is found to have inconsistent $\pi_\e$ from annual parallax and satellite parallax methods, and this inconsistency can be well explained by the full orbital motion of the lens system \citep{Han:2016}. In the case of OGLE-2015-BLG-0196 and OGLE-2016-BLG-0168, the annual parallax effect is confirmed by the satellite parallax method (\citealt{Han:2017}; \citealt{Shin:2017}).

In this paper, we present the analysis of a \emph{Spitzer} binary event MOA-2015-BLG-020. This is the second published case in which the annual parallax effect agrees with the satellite parallax effect. We summarize the ground-based and space-based observations in Section~\ref{sec:observations}, describe the light curve modeling in Section~\ref{sec:modeling}, and derive the physical properties of the binary system in Section~\ref{sec:physics}. In Section~\ref{sec:discussion}, we review all published microlensing binaries in which the annual parallax effect has been confirmed or contradicted by other methods, and discuss the implications.

%%%%%%%%%%%%%%%%%%%%%%%%%%%%%%
\section{Observations} \label{sec:observations}

\subsection{Ground-based Alert and Follow-up}
At UT 10:06 of 2015 February 16 (HJD$'\equiv {\rm HJD}-2450000=7101.87$), the MOA collaboration identified the microlensing event MOA-2015-BLG-020 at equatorial coordinates (R.A., decl.)$_{2000}=(17^{\rm h}52^{\rm m}52\fs78,-32\arcdeg29\arcmin09\farcs1)$, with corresponding Galactic coordinates $(l,b)_{2000}=(-2\fdg24,-3\fdg16)$, based on data taken by its 1.8 m telescope with a 2.2 deg$^2$ field at Mt. John, New Zealand. These MOA observations were taken in a broad $\sim$R+I band pass at 15 minute cadence. 
The OGLE collaboration independently discovered this event about 2.5 days after the MOA alert. It was alerted as OGLE-2015-BLG-0102 through the OGLE Early Warning System \citep{Udalski:1994,Udalski:2003}, based on observations from the 1.4 deg$^2$ camera on its 1.3 m Warsaw Telescope at the Las Campanas Observatory in Chile. This microlensing event, lies in the OGLE-IV field BLG535 \citep{Udalski:2015a}, meaning that it received OGLE observations at a cadence of 2-3 observations per night.

Event MOA-2015-BLG-020 also lies in one of the four prime fields of the Korean Microlensing Telescope Network \citep[KMTNet,][]{Kim:2016}, and thus received dense coverage from KMTNet. In 2015, KMTNet observed a $\sim$16 deg$^2$ prime microlensing fields at $\sim$10 minute cadence when the bulge was visible. The KMTNet consists of three 1.6 m telescopes, with each equipped with a 4 deg$^2$ field-of-view camera. The observations started on February 3, 2015 (HJD$'$=7056.9) for its CTIO telescope, February 19, 2015 (HJD$'$=7072.6) for its SAAO telescope, and June 9, 2015 (HJD$'$=7182.9) for its SSO telescope, respectively.

This event was also observed by the Las Cumbres Observatory Network (LCO, \citealt{Brown:2013}), to support the 2015 \emph{Spitzer} microlensing campaign. See \citet{Street:2016} for more detailed description of LCO observations. Event MOA-2015-BLG-020 received from HJD$'$=7123.7 to 7174.8\ in total 186 observations from two 1-m telescopes at CTIO, 105 observations from two 1-m telescopes at SAAO, and 76 observations from two 1-m telescopes at SSO.

All ground-based data were reduced using the standard or variant version of the image subtraction method developed by \citet{AlardLupton:1998}, employing a spatially variant kernel as necessary (see also \citealt{Bramich:2008}).

\subsection{\emph{Spitzer} Follow-up}
Event MOA-2015-BLG-020 was selected for \emph{Spitzer} IRAC 3.6 $\mu$m observations as part of the 2015 \emph{Spitzer} microlensing campaign to probe the Galactic distribution of planets \citep{SCN:2015a,Zhu:2017}. The general description of the campaign and the target selection protocol can be found in \citet{Udalski:2015b} and \citet{Yee:2015b}, respectively. By the time the 2015 \emph{Spitzer} program started,
\footnote{Although \emph{Spitzer} observations did not start until June 8, 2015 (HJD = 2457182), the target selections started in late May of 2015.}
the binary nature of the current event was already established. Therefore, it was selected as ``subjective binary'' on June 1, 2015 (HJD = 2457175), meaning that \emph{Spitzer} observations were taken specifically for measuring the mass of the binary. Then, observations started on 2015 June 8 (HJD = 2457182), and ended on 2015 July 15 (HJD = 2457219) when this target moved out of \emph{Spitzer}'s Sun-angle window. The cadences were determined objectively, and in total 61 observations were taken.

The \emph{Spitzer} data were reduced by the software that was designed specifically for this microlensing program \citep{SCN:2015b}. In the present case, because the source star is very bright and red, it was saturated on the \emph{Spitzer} images that were taken in the first few days. Although the saturation issue is in principle solvable (e.g., OGLE-2015-BLG-0763, \citealt{Zhu:2016}), we decided to exclude the first 10 data points that are potentially affected by saturation, on the basis that no particularly interesting behavior occurred during this time interval. In the end, we include 51 \emph{Spitzer} observations spanning from HJD$'$=7185.7 to 7221.8 for the parallax measurement.

%%%%%%%%%%%%%%%%%%%%%%%%%%%%%%
\section{Light Curve Modeling} \label{sec:modeling}
\subsection{Initial Solution Search} \label{sec:grid-search}

\begin{figure*}
\centering
\plotone{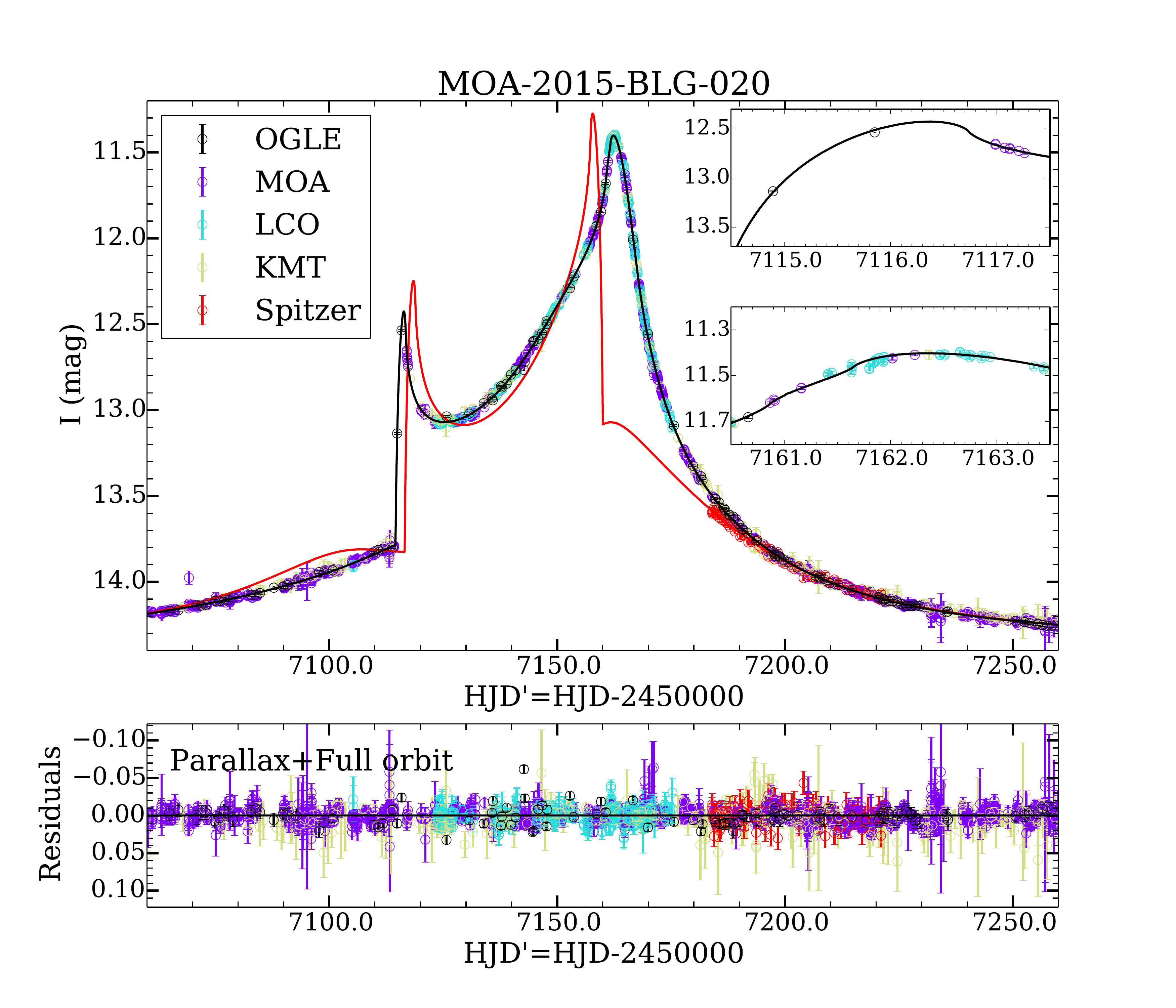}
%\centering\includegraphics[width=1.0\textwidth]{ob150102-lc-1.pdf}
\caption{The light curve of event MOA-2015-BLG-020. In the top panel, the blue and red lines are the best-fit theoretical light curves for ground-based and satellite observations. The bottom panel shows the residual from the best model. Data points from different collaborations are shown with different colors.} \label{fig:lc}
\end{figure*}

The light curve of event MOA-2015-BLG-020 suggests that it is a typical
binary microlensing event (see Figure \ref{fig:lc}). In the standard
terminology (i.e., binary event without parallax and lens orbital motion
effects), the following seven parameters are used for characterizing a
binary light curve: the time of the closest approach between the source
and the binary lens (gravitational) center, $t_0$; the impact parameter
normalized by the Einstein radius, $u_0$; the event timescale, $t_\e$;
the source size normalized by the Einstein radius, $\rho$; the projected
separation between the binary components normalized to the Einstein
radius, $s$; the binary mass ratio, $q$; the angle between the
binary-lens axis and the lens-source relative motion,
$\alpha$. There are two further flux parameters that
  describe the source flux $(F_\s^j)$ and the blending flux $(F_\b^j)$ for each observatory $j$ that translate the magnification $A$ to the observed flux at given time $t_i$
\begin{equation}
    F^j(t_i) = F_\s^j \cdot A(t_i)+F_\b^j\ .
\end{equation}
These flux parameters are found for each data set using linear fit. We use the the advanced contour integration code, \texttt{VBBinaryLensing}
\footnote{\url{http://www.fisica.unisa.it/GravitationAstrophysics/VBBinaryLensing.htm}}
, to compute of the binary lens magnification $A(t_i)$. This code includes a parabolic correction in Green's line integral, and can automatically adjust the step size of integration based on the distance to the binary caustic, in order to achieve a desired precision in magnification. See \citet{Bozza:2010} for more details.

We start with a grid search on the ground-based data alone for the possible binary solution (or solutions). The grid search is conducted on parameters $(\log{s},\log{q},\log{\rho},\alpha)$, with 16 values equally spaced between $-1\le\log{s}\le1$, $-3\le\log{q}\le0$, $-4\le\log{\rho}\le0$, and $0^\circ \le \alpha \le 360^\circ$, respectively. For each set of $(\log{s},\log{q},\log{\rho},\alpha)$, we find the minimum $\chi^2$ by going downhill on the remaining parameters $(t_0,u_0,t_\e)$.

The global minimum is found at $\log{s}\sim0$ ($s\sim1$), $\log{q}\sim-0.6$ ($q\sim0.25$), $\log\,{\rho}\sim-2$ and $\alpha\sim 220$\degree, and there is no other locus on this grid that has similar $\chi^2$.

We then refine the solution by performing Markov Chain Monte Carlo (MCMC) analysis around the initial solution found by the previous grid search, which employs the \texttt{emcee} ensemble sampler \citep{ForemanMackey:2013}.

\subsection{Inclusion of Microlensing Parallax Effect}
The microlensing parallax effect has to be taken into account in order to simultaneously model the ground-based and space-based data. This effect invokes two additional parameters, $\pien$ and $\piee$, which are the northern and eastern components of the parallax vector $\bdv{\pie}$.

We try to constrain $\bdv{\pie}$ based on ground data alone, and by simultaneously modeling ground and \emph{Spitzer} data, in order to check the consistency between annual parallax and satellite parallax effects. This check is possible here because MOA-2015-BLG-020 occurred relatively early in the season and had a fairly long timescale. 

\begin{figure*}
\centering
\epsscale{0.8}
\plotone{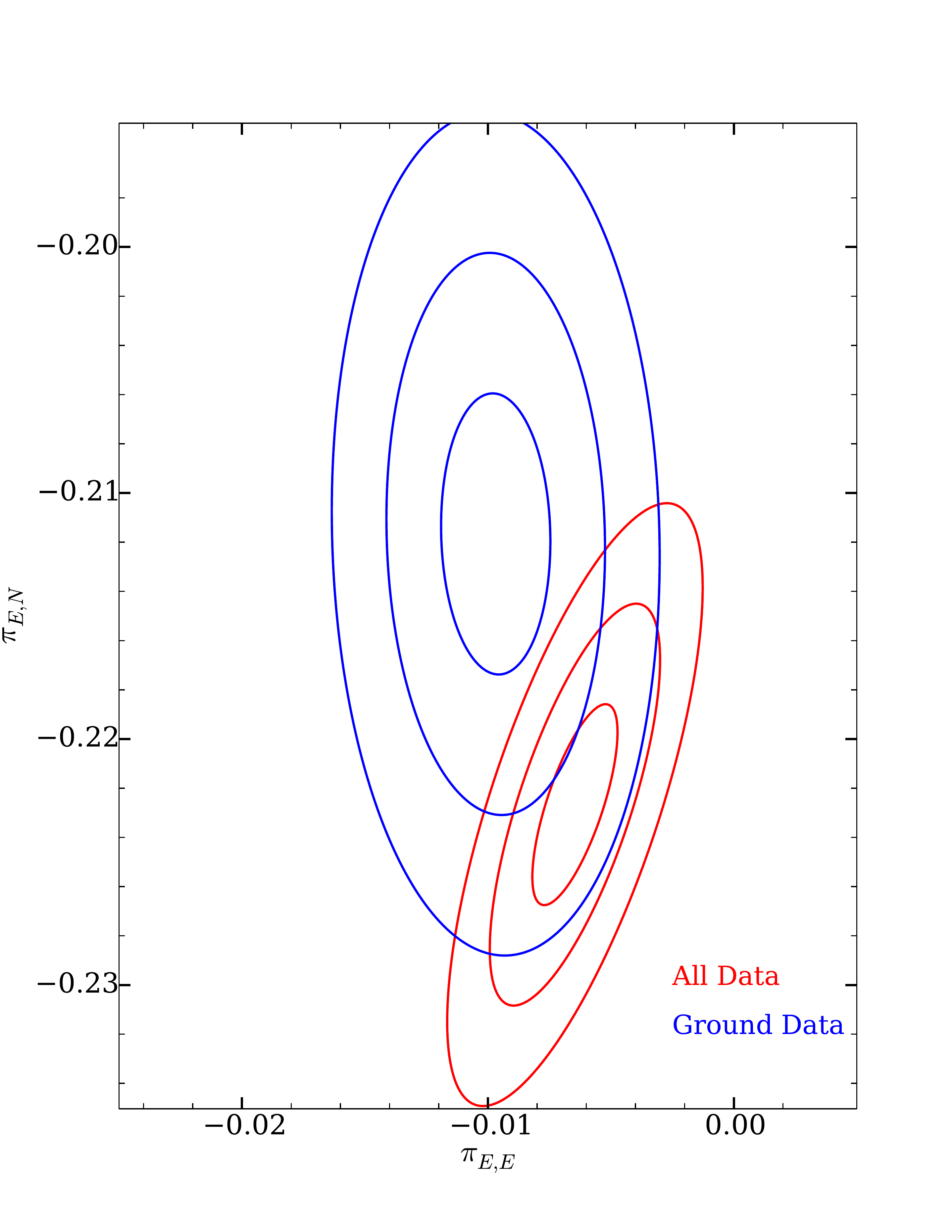}
\caption{Distribution of microlensing parallax parameters $\piee$ and $\pien$ in the East and North directions. The red and blue contours are obtained based on the combined ground+\emph{Spitzer} data and the ground data alone, respectively. The three contours show 1-$\sigma$ ($\chi^2=1$), 2-$\sigma$ ($\chi^2=4$) and 3-$\sigma$ ($\chi^2=9$) confidence regions of the full orbit model.
} \label{fig:parallax}
\end{figure*}

The annual parallax effect leads to two discrete solutions arising from the $\pm u_0$ degeneracy \citep[e.g.,][]{Smith:2003,Poindexter:2005}. In the case of MOA-2015-BLG-020, we find the two solutions have $\Delta\chi^2\geq100$ because of strong annual parallax effect, indicating that the $+u_0$ solution is strongly favored over the $-u_0$ solution. We show in Figure \ref{fig:parallax} the 3-$\sigma$ constraints on $\bdv{\pie}$ based on ground-based data alone. 

We then take into account the satellite parallax effect in order to include \emph{Spitzer} data. We extract the geocentric locations of \emph{Spitzer} during the entire season from the \emph{JPL Horizons} website
\footnote{\url{http://ssd.jpl.nasa.gov/?horizons}}
, and project them onto the observer plane. The projected locations are then oriented and rescaled according to a given $\bdv{\pie}$ to work out \emph{Spitzer}'s view of the microlensing geometry.

We include the $I-[3.6\microm]$ color constraint on the source star to better constrain the parallax parameters, considering the simple monotonic falling behavior of the \emph{Spitzer} light curve. This has been demonstrated to be effective in single-lens cases \citep[e.g.,][]{SCN:2015a,Zhu:2017}. The $I-[3.6\microm]$ color constraint comes from putting the measured $V-I$ source color into the $I-[3.6\microm]$ vs. $V-I$ relation, which is derived based on nearby field stars of similar colors (see \citealt{SCN:2015b} for more details). This yields $I-[3.6\microm]=3.18\pm0.05$ mag. During the MCMC process, we calculate $I-[3.6\microm]$ from the flux parameters (which are found using a linear fit), and %treat $I-[3.6\microm]$ as a free parameter (rather than $F_\s^{\rm spitzer}$), and
reject any values that are $>3\sigma$ away from the central value.

The constraints on $\bdv{\pie}$ from the simultaneous modeling of ground and \emph{Spitzer} data are also shown in Figure \ref{fig:parallax}. We note that these are the constraints from annual parallax and satellite parallax signals together. In order to separate constraints from these two types of parallaxes, we assume the posteriors of the ground-only and ground+\emph{Spitzer} fits are multivariate Gaussians, and derive the covariance matrix of the satellite parallax part (See the Appendix and \citealt{Gould:2003}) The derived matrix has a determinant that is statistically consistent with zero, which indicates a strong correlation between $\pien$ and $\piee$. This is because the \emph{Spitzer} observations sometimes only measure a one-dimensional parallax component (see \citealt{Shvartzvald:2015} for a detailed discussion). We nevertheless proceed with the standard procedure and compute the difference between the annual parallax and the satellite parallax measurements, and find $\Delta \chi^2=11$. Although this would formally indicate a probability of $e^{-11/2}=0.4\%$, it is actually well within the systematic uncertainty that the ground-based data can introduce.
\footnote{In other words, we would never consider it a reliable parallax measurement if the annual parallax only has $\Delta\chi^2=11$ improvement compared to the standard model.}
%The overlap between the two constraints suggests consistency between annual parallax and satellite parallax effects. Figure \ref{fig:parallax} only shows the ground-satellite parallax consistency in full orbit model, but in fact $\pie$ in all models are consistent with each other (see Table \ref{table1}). We also calculate the difference between the satellite and annual parallax using covariance matrix, and the result is $\chi^2=11$. Although this would normally suggest a 3.3-sigma inconsistency, it is significantly smaller than the $\chi^2$ improvement (5907-4162=1745) that the annual parallax signal introduces in the ground-base data. 
Therefore, the small $\chi^2$ indicates the good agreement between the annual parallax and the satellite parallax.  

Notice that we took into account the four-fold parallax degeneracy (which are often denoted as $(++)$, $(+,-)$, $(-,+)$ and $(-,-)$). Two of these are eliminated since the $-u_0$ solution is excluded by the ground-based data, while the $(+,-)$ degeneracy from Spitzer is also eliminated. 

\subsection{Inclusion of Binary Lens Orbital Motion Effect}

We then introduce the lens orbital motion effect into the light curve modeling. Despite the degeneracy between orbital motion and parallax, as we will see below, in this case, this has no effect on the measured values or uncertainties of the parallax. We introduce orbital motion into the models in two different ways. First, we use the linear orbital motion approximation, which involves two parameters $\dif\alpha/\dif t$ and $\dif s/\dif t$. For the binary system to remain bound, we also impose the constraints on the projected kinetic to potential energy ratio \citep{Dong:2009}. Second, we also include $z$ and $\dif z/\dif t$ in addition to $\dif\alpha/\dif t$ and $\dif s/\dif t$, in order to account for the full Keplerian motion of the binary system. Here $z$ and $\dif z/\dif t$ quantify the binary separation (normalized to the Einstein radius) along the line of sight and its time derivative, respectively. See \citet{Skowron:2011} for the conversion between these phase-space parameters and Keplerian parameters. This conversion requires an input of the source angular size $\theta_\star$ in order to set the absolute physical scale, and we use $\theta_\star=23.9~\microas$ (see Section~\ref{sec:physics}). Once the Keplerian parameters are derived, we check the orbital period $P$ and reject any solution with $P_{\rm orb}\ge 200$~yrs, in order to avoid the influence of systematics in the data.

The results of two modelings with different treatment of the orbital motion are given in Table~\ref{table1}; the source trajectories and caustics of the full orbit motion are shown in Fig. \ref{fig:caustics}. This solution has slightly worse $\chi^2$ than the linear orbital motion solution, even though the former has two more free parameters. This is because the linear orbit solution (with the ratio of the perpendicular kinetic energy to the potential energy, $\beta = {KE_{\rm perp}}/{PE} = 0.145$) has preferentially long orbital periods ($P_{\rm orb}\gg200$~yrs) for the binary system, which are not allowed in the full orbit solution. Nevertheless, the microlensing parameters (especially the microlens parallax) are stable regardless of whether and how the lens orbital motion is included. 

%With the lens properties we derive in the next section, we fit the data to a full orbital motion model. This model has a $\chi^2$ a bit larger than the linear orbit model because the linear model, when transferring to a full-orbit one, would have a several-million-year period, which is unphysical. So we set a period constraint $T<200$yr which only influences 8\% samples in the MCMC chain but prevents those solutions with an unphysical long period. We will return to this briefly in the discussion.

%%%%%%%%%%%%%%%%%%%%%%%%%%%%%%
\section{Physical Parameters} \label{sec:physics}

\begin{figure*}
\centering
\plotone{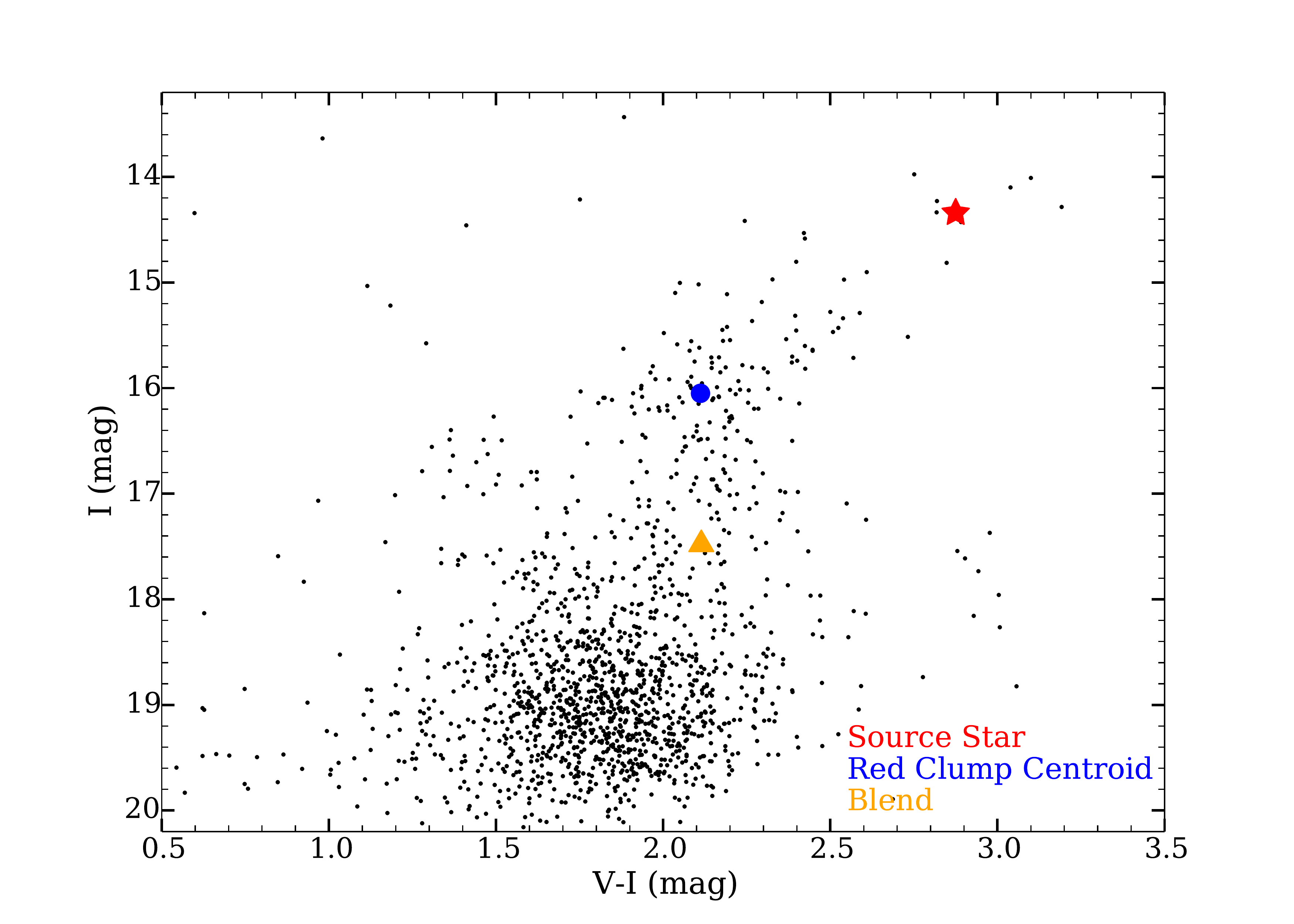}
\caption{OGLE-IV calibrated color magnitude diagram of the stars (black dots) within 2'$\times$2' of MOA-2015-BLG-020/OGLE-2015-BLG-0102. The blue dot shows the centroid of the red clump stars. The red asterisk indicates the position of the microlensed source, and the yellow triangle shows the position of the blended object.} \label{fig:cmd}
\end{figure*}

We estimate the angular size of the source following the standard procedure \citep{Yoo:2004}. First, we measure the centroid of the red clump in the OGLE $(V-I,I)$ color magnitude diagram (CMD) of the stars within 2'$\times$2' of our event (see Fig.~\ref{fig:cmd}). By using stars in the box $1.8<V-I<2.4$ and $15.5<I<16.5$, we find the centroid of the red clump to be $(V-I,I)_{\rm RC}=(2.11\pm0.05,16.05\pm0.11)$. This, when combined with the instrumental color and magnitude of the source star $(V-I,I)_{\rm S, OGLE}=(2.87,14.34)$, yields an offset of $\Delta(V-I,I)_{\rm OGLE}=(V-I,I)_{\rm S, OGLE}-(V-I,I)_{\rm RC,OGLE}=(0.76, -1.71)$. After applying the correction of the non-standard $V$ band of OGLE-IV \citep{Udalski:2015a,Zhu:2015}:$\Delta(V-I)_{\rm JC}=\Delta(V-I)_{\rm OGLE}\times0.92=0.70$ (in which ``JC'' represents the standard Johnson-Cousins system) and adapting the intrinsic color and magnitude of the clump $(V-I,I)_{\rm RC,0}=(1.06,14.56)$ \citep{Bensby:2013,Nataf:2013}, we find that the intrinsic color and magnitude of source is $(V-I,I)_{\rm S,0}=\Delta(V-I, I)_{\rm JC}+(V-I, I)_{\rm RC,0}=(1.76, 12.85)$. 

To determine the source angular size, we employ the color-surface brightness relation of giant stars from \citet{Kervella:2004}, and finally find
\begin{equation}
\theta_\star = 23.9\pm1.0\  \microas.
\end{equation}
Therefore,
\begin{equation}
\theta_E = 1.329\pm0.049\  \mas.
\end{equation}
Then we obtain the total mass of the system using equation \ref{equ:mass}:
\begin{equation}
M = 0.731\pm0.034 M_\odot.
\end{equation} 
Combined with the mass ratio from our fit, we find out that the lens system is a binary of $0.606M_\odot$ and $0.125M_\odot$. The lens-source relative parallax is $\pi_\rel=0.296\pm0.017~\mas$, indicating a disk binary. Under the assumption that the source is at the same distance as the red clump centroid at this location ($D_{\rm S}= 8.8~\text{kpc}$, \citealt{Nataf:2013}), the distance to the lens is
\begin{equation}
D_{\rm L} = 2.44\pm0.10\ \text{kpc},
\end{equation} 
The lensing binary is vertically about 120\,pc away from the Galactic plane, likely from the thin disc.

The binary components are separated in projection by 
\begin{equation}
r_\perp = 4.04\pm0.23\ \au.
\end{equation} 
We summarize the physical parameters in Table \ref{phys}.

%%%%%%%%%%%%%%%%%%%%%%%%%%%%%%
\section{Discussion} \label{sec:discussion}
We analyzed the binary-lensing event MOA-2015-BLG-020 which was observed both from the ground and from the \emph{Spitzer Space Telescope}. The light curve from ground-based observations significantly deviated from the lensing model based on the rectilinear lens-source relative motion and we measure the microlensing parallax from the analysis of the deviation. The measured parallax was confirmed by the \emph{Spitzer} data, showing the consistency between annual parallax and satellite parallax effects. By additionally measuring the angular Einstein radius from the analysis of the resolved caustic crossing, the mass and distance to the lens are determined. We find that the lens is a binary composed of two low-mass stars located in the Galactic disk. In our analysis, we find that the linear orbit model and full orbit model can fit the data almost equally well. Although the full orbit parameters $z$ and $\dif z/\dif t$ are not well constrained, we report lens physical parameters based on the full orbit model because of its physical foundation \citep{Han:2016}. 

\begin{figure*}
\centering
\plotone{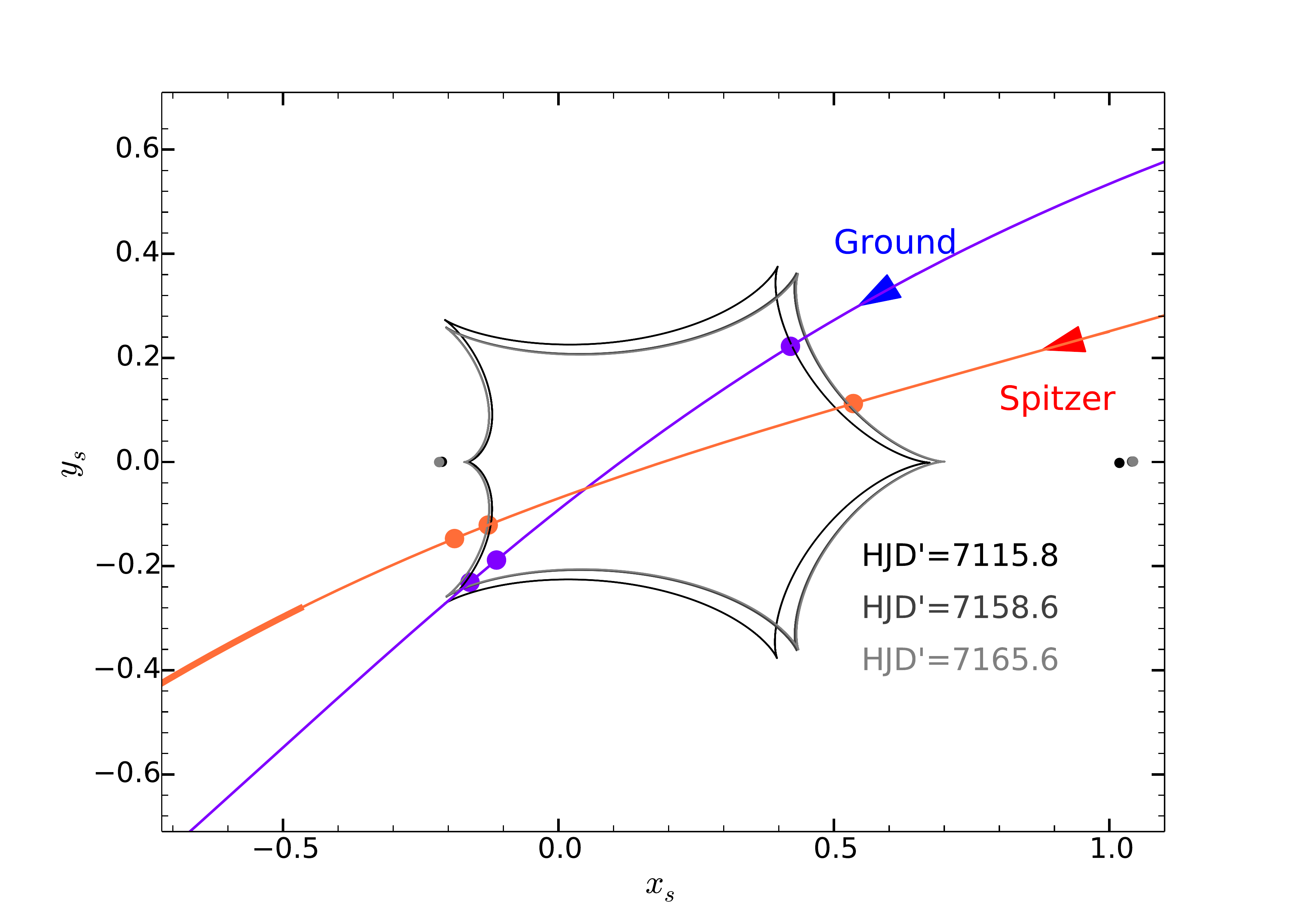}
\caption{Caustics patterns for the event MOA-2015-BLG-020. Three caustic curves are shown for three different epochs (caustic entrance from the ground, caustic exits from \emph{Spitzer} and ground), although the two at HJD'=7158.6 and 7165.6 almost overlap. The corresponding lens and source positions are shown as solid dots. The source trajectories for the ground and Spitzer are shown as blue and red curves respectively. The arrows indicate the directions of the source motions. The bold line segment indicates the epochs of the Spitzer data. } \label{fig:caustics}
\end{figure*}

The binary-lensing event MOA-2015-BLG-020 is peculiar in one aspect. Usually for a caustic-crossing binary event, the light curve has one sharp rise and one sharp fall during the caustic entrance and exit, respectively, producing a `U'-shaped light curve. However, for this binary event, the caustic exit gracefully merges with the cusp crossing (see Fig.~\ref{fig:caustics}), which causes the absence of the sharp decline in the ground-based light curve. This is almost a direct result of lens orbital motion. Note that this sharp decline is predicted to be present in the space-based light curve, but it is not observable by \emph{Spitzer} due to various observational constraints (see \citealt{Udalski:2015b} for details).
 
In Table \ref{tab:LPer}, we summarize the published microlensing binaries in which the parallax parameters detected from ground (through annual parallax effect) are tested by other methods. The parallax parameters based on ground data in OGLE-2011-BLG-0417 and OGLE-2015-BLG-0479 seem to be inconsistent with the results of independent checks. Although there is still possibility that the parallax detected by microlensing is wrong, it is believed that the radial velocity measurements for OGLE-2011-BLG-0417 do not test the parallax model. The reason is that the blended light of OGLE-2011-BLG-0417 was not the lens because it is brighter than predicted flux from lens. The inconsistency in OGLE-2015-BLG-0479 is expected because of the strong lens orbital motion effect. Indeed, it is because of this inconsistency that the full Keplerian parameters of the binary in OGLE-2015-BLG-0479 can be well constrained \citep{Han:2016}.

%We still do not know how this discrepancy arises for OGLE-2011-BLG-0417, but for OGLE-2015-BLG-0479, it is caused by the unphysical nature of the linear-orbit model \citep{Han:2016}. This is why we attempted a full-orbit model instead of a linear-orbit model in our work, even though the full-orbit model is not well constrained but it gives a reasonable period of about $70\,{\rm yr}$.

For the remaining events, the microlensing parallax parameters from annual parallax effect are confirmed by additional observations (high-resolution imaging, radial velocity, or satellite parallax). As expected, they all have relatively long timescales ($t_\e\gtrsim60~$days), and the majority of them peaked (as seen from ground) either early (before May) or late (after August) in the microlensing season.
Among these events, half are confirmed by ground observations (by adaptive optics or radial velocity) and they all have a small impact parameter $u_0$, while those confirmed by satellite observations could have larger $u_0$ (modest magnification). This highlights \emph{Spitzer}'s power to discover parallax events for moderately magnified events. The abundance of stellar binaries with $-1 \leq \log q \leq 0$ is roughly uniform as a function of $\log q$ ($\dif N/\dif q \propto q^{-1}$), consistent with \cite{Trimble:1990}, although we caution that the number of events is small and no selection effects have been taken into account. As more and more lenses with definite masses are determined from microlensing, it will be very interesting to study this statistics much more carefully in the future.

%%%%
\acknowledgements
This work has been supported in part by the National Natural Science Foundation of China (NSFC) grants 11333003 and 11390372 (SM). This research uses data obtained through the Telescope Access Program (TAP), which has been funded by the Strategic Priority Research Program ``The Emergence of Cosmological Structures" (Grant No. XDB09000000), National Astronomical Observatories, Chinese Academy of Sciences, and the Special Fund for Astronomy from the Ministry of Finance. Work by WZ, YKJ, IGS and AG was supported by AST-1516842 from the US NSF and JPL grant 1500811. Work by JCY was performed in part under contract with the California Institute of Technology (Caltech)/Jet Propulsion Laboratory(JPL) funded by NASA through the Sagan Fellowship Program executed by the NASA Exoplanet Science Institute. This research has made use of the KMTNet system operated by the Korea Astronomy and Space Science Institute (KASI) and the data were obtained at three host sites of CTIO in Chile, SAAO in South Africa, and SSO in Australia. The OGLE Team thanks Drs. G. Pietrzy{\'n}ski and {\L}. Wyrzykowski for their contribution to the collection of the OGLE photometric data over the past years. The OGLE project has received funding from the National Science Centre, Poland, grant MAESTRO 2014/14/A/ST9/00121 to AU. Work by C.H. was supported by the grant (2017R1A4A1015178) of National Research Foundation of Korea. This work makes use of observations from the LCO network, which includes three SUPAscopes owned by the University of St Andrews. The RoboNet programme was an LCO Key Project using time allocations from the University of St Andrews, LCOGT and the University of Heidelberg together with time on the Liverpool Telescope through the Science and Technology Facilities Council (STFC), UK. This research has made use of the LCO Archive, which was operated by the California Institute of Technology, under contract with the Las Cumbres Observatory. The MOA project is supported by JSPS KAKENHI Grant Number JSPS24253004, JSPS26247023, JSPS23340064, JSPS15H00781, and JP16H06287.

\software{VBBinaryLensing(\url{http://www.fisica.unisa.it/GravitationAstrophysics/VBBinaryLensing.htm}), emcee\citep{ForemanMackey:2013}}

\appendix 
Let $a$ and $c$ denote a parallax measurement and its covariance matrix, and the subscripts ``comb'' and ``ground'' indicate quantities appropriate for the combined and ground-based data. For example, $a_{\rm comb}$ and $c_{\rm comb}$ are the measured parallax and its covariance matrix from our MCMC fits for the combined data.  We follow \citet{Gould:2003} to derive the ``Spitzer'' parallax and the consistency between the ground and Spitzer parallaxes (expressed as $\Delta \chi^2$) through the following steps. 
First, we  introduce several quantities for later use
\begin{equation*}
\begin{split}
&b_{\rm comb} = (c_{\rm comb})^{-1}, b_{\rm ground} = (c_{\rm ground})^{-1},\\
&(d_{\rm comb})_i = \sum_{j}(b_{\rm comb})_{ij}\times(a_{\rm comb})_j,\\
&(d_{\rm ground})_i = \sum_{j}(b_{\rm ground})_{ij}\times (a_{\rm ground})_j.\\
\end{split}
\end{equation*}   
Then we calculate the corresponding quantities for the Spitzer parallax
\begin{equation*}
\begin{split}
&b_{\rm spitzer} = b_{\rm comb}-b_{\rm ground}, d_{\rm spitzer}=d_{\rm comb}-d_{\rm ground},\\
&c_{\rm spitzer} = (b_{\rm spitzer})^{-1},\\
&(a_{\rm spitzer})_i = \sum_{j}(c_{\rm spitzer})_{ij}\times (d_{\rm spitzer})_j.\\
\end{split}
\end{equation*}
Finally we compute the difference between the annual  and satellite parallax measurements: 
\begin{equation*}
\Delta\chi^2 = \sum_{ij}(a_{\rm diff})_i\times (b_{\rm diff})_{ij}\times (a_{\rm diff})_j,
\end{equation*}
where
\begin{equation*}
\begin{split}
&a_{\rm diff} = a_{\rm ground} - a_{\rm spitzer},\\
&c_{\rm diff} = c_{\rm ground} + c_{\rm spitzer},\\
&b_{\rm diff} = (c_{\rm diff})^{-1}.\\
\end{split}
\end{equation*}

\clearpage
\begin{deluxetable*}{lccccccc}
\centering

\tablecaption{Best-fit parameters.
\label{table1}}
\tablehead{\colhead{Parameters$^1$} & \multicolumn{3}{c}{Ground+\emph{Spitzer}} & \multicolumn{4}{c}{Ground-only} \\
\colhead{} & \colhead{Full Orbit} & \colhead{Linear Orbit} & \colhead{Parallax Only} & \colhead{Full Orbit} & \colhead{Linear Orbit} & \colhead{Parallax Only} &\colhead{Without Parallax}}
\startdata
$\chi^2/$dof	&4003.5 / 4028	&4000.8 / 4030	&4222.4 / 4032	&3940.1 / 3967	&3936.4 / 3969	&4162.0 / 3971	&5907.2/3973\\
$\log{s}$    	&$0.0962\pm0.0003$	&$0.0963\pm0.0003$		&$0.0940\pm0.0002$	&$0.0962\pm0.0003$	&$0.0963\pm0.0003$	&$0.0940\pm0.0002$	&$0.08767\pm0.00009$	\\
$\log{q}$    	&$-0.684\pm0.002$	&$-0.683\pm0.002$		&$-0.703\pm0.001$	&$-0.685\pm0.002$	&$-0.685\pm0.002$	&$-0.703\pm0.001$	&$-0.7477\pm0.0007$	\\
$u_0$ 	    	&$0.0716\pm0.0008$	&$0.0715\pm0.0008$		&$0.0798\pm0.0004$	&$0.0711\pm0.0008$	&$0.0715\pm0.0008$	&$0.0799\pm0.0004$	&$0.09109\pm0.00015$	\\
$t_0$ 	    	&$7145.48\pm0.03$	&$7145.46\pm0.03$		&$7145.72\pm0.03$	&$7145.48\pm0.03$	&$7145.51\pm0.03$	&$7145.71\pm0.03$	&$7145.66\pm0.01$	\\
$t_E$ (days)   	&$63.56\pm0.06$		&$63.55\pm0.06$			&$62.94\pm0.06$		&$63.74\pm0.07$		&$63.61\pm0.08$		&$62.94\pm0.06$		&$64.69\pm0.03$	\\
$\log{\rho}$  	&$-1.745\pm0.002$	&$-1.744\pm0.002$		&$-1.757\pm0.002$	&$-1.746\pm0.002$	&$-1.745\pm0.002$	&$-1.757\pm0.002$	&$-1.7969\pm0.0009$	\\
$\alpha$ (deg)	&$219.12\pm0.08$	&$219.14\pm0.08$		&$218.27\pm0.05$	&$219.18\pm0.08$	&$219.14\pm0.09$	&$218.28\pm0.05$		&$217.43\pm0.03$\\
$\pien$	    	&$-0.223\pm0.004$	&$-0.223\pm0.005$		&$-0.220\pm0.006$	&$-0.212\pm0.006$	&$-0.211\pm0.010$	&$-0.218\pm0.006$	&\nodata	\\
$\piee$	    	&$-0.007\pm0.002$	&$-0.006\pm0.002$		&$0.003\pm0.002$	&$-0.010\pm0.002$	&$-0.010\pm0.002$	&$0.003\pm0.002$		&\nodata\\
$\frac{ds}{dt}$	(year$^{-1}$) &$0.212\pm0.015$	&$0.217\pm0.017$		&\nodata		&$0.240\pm0.017$	&$0.225\pm0.018$	&\nodata		&\nodata\\
$\frac{d\alpha}{dt}$	(rad/year) &$0.383\pm0.014$&$-1.197\pm0.038$		&\nodata		&$0.371\pm0.017$	&$-1.197\pm0.040$	&\nodata		&\nodata\\
$z$		&$7.03\pm3.17$		&\nodata			&\nodata		&$4.78\pm4.52$		&\nodata		&\nodata		&\nodata\\
$\frac{dz}{dt}$	(year$^{-1}$)&$0.037\pm0.258$	&\nodata			&\nodata		&$0.277\pm0.270$	&\nodata		&\nodata 	&\nodata		\\
$V-I$		&$2.875\pm0.003$	&$2.874\pm0.003$		&$2.867\pm0.003$	&$2.876\pm0.003$	&$2.874\pm0.003$			&$2.868\pm0.002$		&$2.860\pm0.002$\\
\enddata
\tablecomments{$^1$ We use $\theta_*=23.9~\microas$ in our full orbit model.}
\end{deluxetable*}

\begin{deluxetable}{lc}
\tablecaption{Physical parameters of the binary lens system MOA-2015-BLG-020. \label{phys}}
\tablehead{\colhead{Parameters} & \colhead{}}
\startdata
	$\theta_E$  (mas)				&1.329$\pm$0.049\\
	$\pi_{rel}$ (mas)				&0.296$\pm$0.017\\
	$M_1$ ($M_\odot$)			&0.606$\pm$0.028	\\
	$M_2$ ($M_\odot$)			&0.125$\pm$0.006	\\
	Distance to lens (kpc)			&2.44$\pm$0.10	\\
	Projected separation (AU)		&4.04$\pm$0.23	\\
	Geocentric proper motion (mas ${\rm yr}^{-1}$)		&7.64$\pm$0.28	\\
\enddata   
\end{deluxetable}

\begin{sidewaystable*}
\caption{Summary of events that have annual parallax tested by other measurements.}\label{tab:LPer}
%\centering 
\tiny{
\begin{tabular}{lcccccccccc}  
	\tableline\tableline
        Event Name			&Confirmed? 		&($\pien$,$\piee$)							&   Mass Ratio(s)		&$t_\e$ (day)			&$t_0$(HJD$'$)		&$u_0$		& $I_\s$	&$\Delta\chi^2_{\rm parallax}$	&$\Delta\chi^2_{\rm orbit}$	&References\\
	\tableline
	\textbf{High Resolution Imaging:}\\
        OGLE-2005-BLG-071	&Yes	&($-0.30^{+0.24}_{-0.28}$, $-0.26\pm0.05$)		&$1.3\times10^{-4}$		&71.1				&3480.7 (04/20)		&0.0282	&19.5							&$<100$			&\nodata			& 1, 2	\\
        OGLE-2006-BLG-109	&Yes	&($-0.316\pm0.013$, $0.139\pm0.006$)	&$(1.4,0.5)\times10^{-3}$						&128.1	&3831.0 (04/05) &0.00345 &20.9 &13.5	&\nodata &3, 4\\
        OGLE-2007-BLG-349 & Yes &$0.204\pm0.034$\tablenotemark{a} & $(3.4\times10^{-4},0.49)$ & 118 & 4348.7 (09/05) & $-0.00198$ & 20.4 & 153 & \nodata & 5 \\
	\textbf{Radial Velocity:}\\
        OGLE-2009-BLG-020	&Yes	&($-0.022\pm0.086$, $0.149\pm0.010$)	&0.273	&76.9				&4917.3 (03/26)		&0.0613	&16.4	&\nodata			&26.3			& 6, 7	\\
	OGLE-2011-BLG-0417	&No		&($0.375\pm0.015$, $-0.133\pm0.003$)	&0.292	&92.3				&5813.3 (09/08)	&-0.0992	&\nodata	&2024	&656	& 8, 9, 10\\
	\textbf{Satellite Parallax\tablenotemark{b}:}\\
        OGLE-2014-BLG-0124	&Yes	&($0.018\pm0.012$, $0.108\pm0.023$)	&$0.752\times10^{-3}$	&131			&6836.2 (06/27)		&0.2099	&18.6
					&\nodata			&\nodata			& 11					\\
        OGLE-2015-BLG-0479\tablenotemark{c}&No\tablenotemark{d}	&($-0.06\pm0.01$, $-0.11\pm0.01$)		&0.81	&86.3				&7166.4 (05/23)		&0.417	&19.6						&\nodata			&43.5			& 12\\
	OGLE-2015-BLG-0196	&Yes	&($0.198\pm0.016$, $0.100\pm0.009$)	&0.99	&96.7				&7115.5 (04/03)		&-0.037	&15.9	&2527.6			&24.4			& 13\\
	OGLE-2016-BLG-0168     &Yes		&($0.382\pm0.022$,$0.057\pm0.011$)	&0724	&97.0				&7492.5 (04/13)		&-0.201	&19.5	&159.4			&115.04		&14\\
	MOA-2015-BLG-020	&Yes	&($-0.223\pm0.005$, $-0.007\pm0.002$)	&0.207	&63.6				&7145.5 (05/03)		&0.0716		&14.3					&1983.5			&218.9			& This work \\				
\tableline
\end{tabular}
}
\tablenotetext{a}{\citet{Bennett:2016} used a different parameterization, with which the amplitude of the parallax vector was constrained.}
\tablenotetext{b}{The mirolens parameters are all from ground-based detection unless otherwise stated. The color, magnitude and $\Delta \chi^2$ are from the combination of ground and satellite data sets.}
\tablenotetext{c}{Parameters of OGLE-2015-BLG-0479 are from the analysis of the combination data set.}
\tablenotetext{d}{It is noticeable based purely on the ground-based data that the full orbital motion is detectable in this event. Therefore, this inconsistency is expected.}
\tablenotetext{}{References: 1. \citet{Udalski:2005}; 2. \citet{Dong:2009}; 3. \citet{Gaudi:2008}; 4. \citet{Bennett:2010}; 5. \citet{Bennett:2016}; 6. \citet{Skowron:2011}; 7. \citet{Yee:2016}; 8. \citet{Shin:2012}; 9. \citet{Boisse:2015}; 10. \citet{Santerne:2016}; 11. \citet{Udalski:2015b}; 12. \citet{Han:2016}; 13. \citet{Han:2017}; 14. \citet{Shin:2017}.}
\end{sidewaystable*}

%\clearpage

\begin{figure*}
\centering
%\plotone{triangle-full.pdf}
\includegraphics[width=1.0\textwidth]{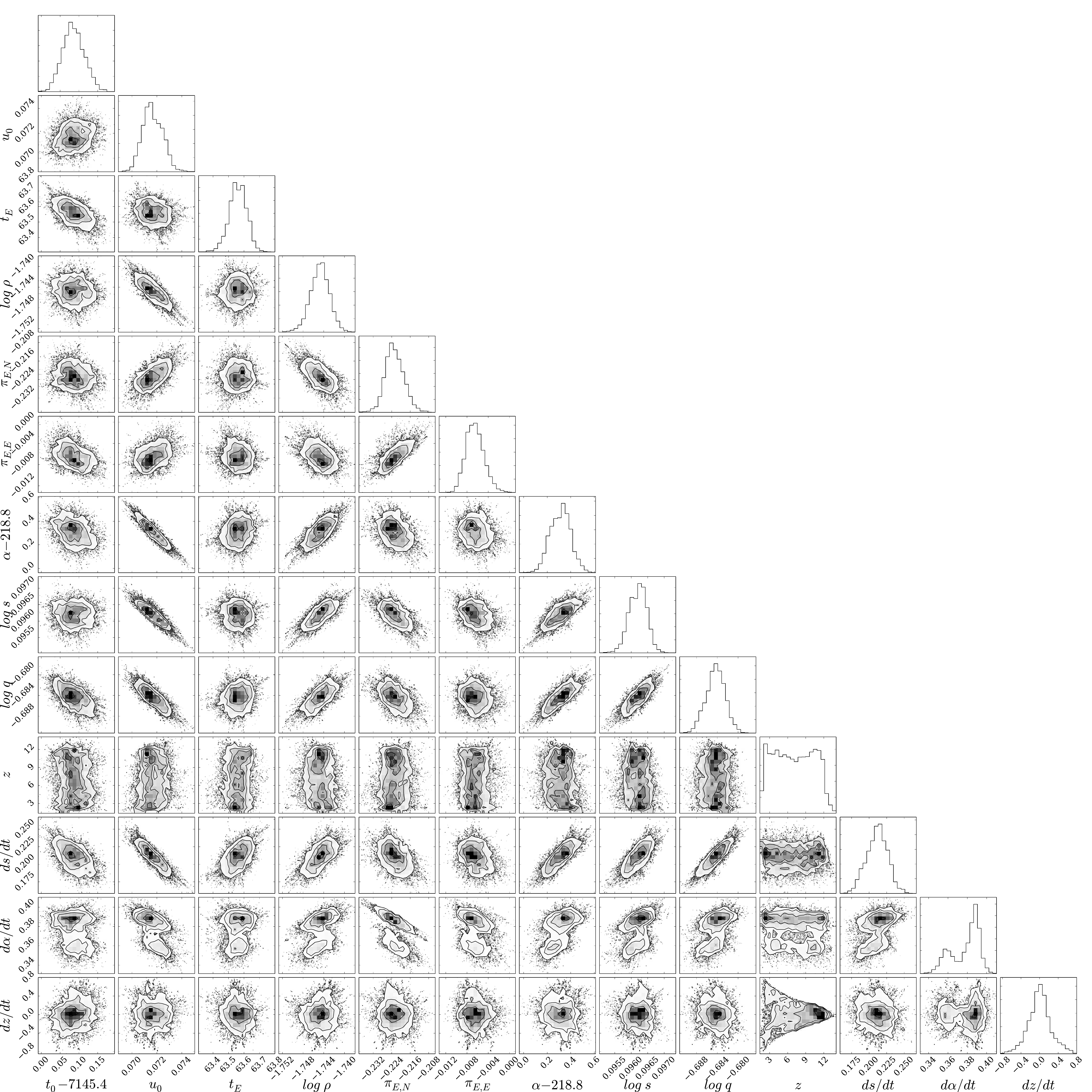}
\caption{The triangle plot of the full orbit model with both ground-based and space-based data included in the fit. The orbital parameters (especially $z$) are not constrained as well as the standard microlensing parameters. Notice that the bimodal distribution of $\frac{d\alpha}{dt}$ is caused by excluding samples with very long period (millions of years). Without such an exclusion, there would be only one peak.
\label{fig:figure4}}
\end{figure*}

\end{CJK*}
\end{document}